\documentclass[arabic, epsf,twocolumn,showpacs,floatfix,preprintnumbers,amsmath,amssymb]{revtex4}
\usepackage{graphicx}
\input{epsf}
\epsfclipon

\begin{document} 

\title{Statistical properties of derivatives: a journey in term structures}

\author{Delphine Lautier}
\affiliation{University Paris-Dauphine, DRM-Finance UMR CNRS 7088, \\Place du Mar\'echal de Lattre de Tassigny
75775 PARIS Cedex 16}
\affiliation{Laboratoire de Finance des March\'es de l'\'Energie (CREST, University Paris-Dauphine, Polytechnique, EDF)}

\author{Franck Raynaud}
\affiliation{University Paris-Dauphine, DRM-Finance UMR CNRS 7088, \\Place du Mar\'echal de Lattre de Tassigny
75775 PARIS Cedex 16}

%\date{\today}

\begin{abstract}
This article presents an empirical study of thirteen derivative markets for commodity and financial assets. It compares the statistical properties of futures contracts's daily returns at different maturities, from $1998$ to $2010$ and for delivery dates up to $120$ months. The analysis of the fourth first moments of the distribution shows that the mean and variance of the commodities follow a scaling behavior in the maturity dimension. The comparison of the tails of the probability distribution according to the expiration dates also shows that there is a segmentation in the \emph{fat} tails exponent term structure above the L\'evy stable region. Finally, the test of the robustness of the inverse cubic law in the maturity dimension shows that there are two regimes of extreme events for derivative markets, reminding of a phase diagram with a transition value at the $18$th delivery month.
\end{abstract}

\maketitle
%%%%%%%%%%%%%%%%%%%%%%%%%%%%%%%%%%%%%%%%%%%%%%%%%%%%%%%%%%%%%%%%%%%%%%%%%%%%%%%%
\section{Introduction}\label{intro}
%%%%%%%%%%%%%%%%%%%%%%%%%%%%%%%%%%%%%%%%%%%%%%%%%%%%%%%%%%%%%%%%%%%%%%%%%%%%%%%%
In the past twenty years, physicists have made several investigations in the fields of social sciences and economics. Their interest in economic systems was risen from the strong analogy between financial markets and complex systems. Both indeed are open systems, far from equilibrium, with macroscopic properties emerging from sub-units interacting non trivially. Therefore, numerous concepts and methods, such as scaling, universality, chaos, agent-based models, have been successfully used to perform empirical investigations and develop financial markets' modeling (\cite{Mantegna1995}, \cite{Mantegna1996}, \cite{Lillo2003}, \cite{Chakraborti2010}).\\
Among all studies, several addressed the question of statistical properties of market prices' fluctuations. It has been shown that irrespective of the particular asset under consideration, prices' fluctuations distribution is characterized by a fat tail with an exponent close to $3$ (\cite{Gopi1999}, \cite{Lux1996}, \cite{Gopi1998}). The majority of these studies provide results for stocks or indexes but there is a lack of information about commodities in the econophysics literature. Moreover, as these studies mainly deal with spot prices, an important and non trivial temporal aspect of derivative markets is missing: the investigation of futures prices.\\
Commodity markets have experienced important evolutions in the last decades: high volatility in the prices, rise in transaction volumes, stronger presence of financial investors seeking for diversification. The introduction of futures contracts with longer delivery dates accompanied this evolution. It confirmed the necessity to understand and manage the term structure of commodity prices, that is to say the relationship at a date $t$, between futures contract having different maturities $M$. Thus term structure models for commodity prices have been developed and improved (\cite{Brennan1985}, \cite{Schwartz1997}, \cite{Cortazar2003}, among others). Inspired by contingent claim valuation models previously built for interest rates (\cite{Vasicek1977}), they were essentially gaussian. Such developments induce two questions. First, are gaussian assumptions suited for commodity prices, especially when long-term delivery dates are concerned? In other words, do the short- and long-term futures prices behave alike? Second, do commodities behave like other derivative assets? \\
This article aims at answering these questions. The statistical properties and characteristics of commodity prices having long-term delivery dates indeed have been relatively few explored. Moreover, a comparison in the maturity dimension with other assets is missing.\\
This article proceeds as follows. After the presentation of the markets and data selected for the study, we analyze the statistical properties of futures prices and in the last section we characterize the tails of the distribution.

%%%%%%%%%%%%%%%%%%%%%%%%%%%%%%%%%%%%%%%%%%%%%%%%%%%%%%%%%%%%%%%%%%%%%%%%%%%%%%%%
\section{Empirical Data}\label{EmpData}
%%%%%%%%%%%%%%%%%%%%%%%%%%%%%%%%%%%%%%%%%%%%%%%%%%%%%%%%%%%%%%%%%%%%%%%%%%%%%%%%

\begin{figure}
  \includegraphics[width=8.6cm,clip]{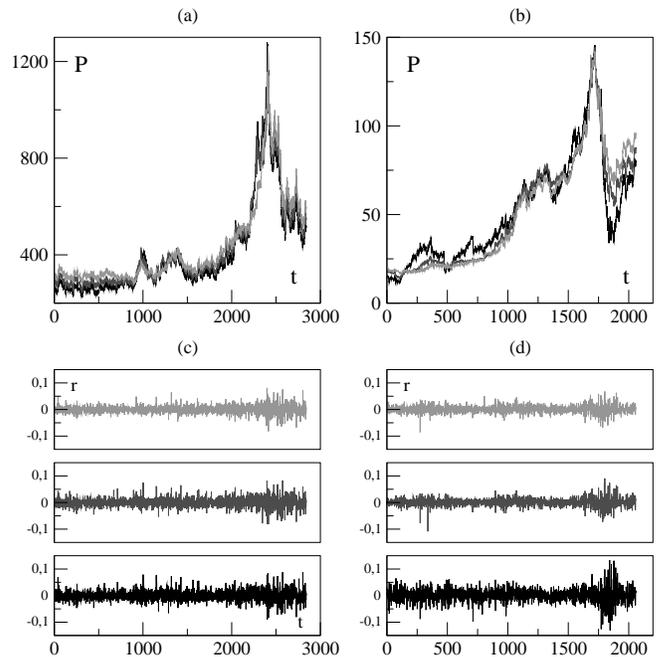}
\caption{Wheat ($C=0.88$) and Light crude oil futures ($C=0.43$) prices and returns corresponding to different delivery dates, $1998$-$2009$. (a) Wheat prices for the maturities: $3$ months (dark), $7$ months (dark gray) and $15$ months (gray); (b) Light crude oil prices for the maturities: $1$ (dark) $24$ (dark gray) and $84$ months (gray); (c) and (d) corresponding daily returns with maturities increasing from the bottom to the top. Time is given in records.}
\label{ret_price}      
\end{figure}

%%%%%%%%%%%%%%%%%%%%%%%%%% Table commodities %%%%%%%%%%%%%%%%%%%%%%%%%
\begin{table}[!t]
\caption{Main characteristics of the collected data: Nature of the assets, trading location, time period, last available maturities, number of records and the contango index C.\\}
\label{tab_commo}
\resizebox{0.475\textwidth}{!}{
\begin{tabular}{|l||l|l|l|l|l|}
\hline
Underlying asset & Exchange-Zone & Period & Maturities & Records & $C$\\
\hline
Light crude oil & CME-US & 1998-2009 & up to 84 &2965 & 0.43\\
\hline
Brent crude & ICE-Eu & 2000-2009 & up to 18 &2523 & 0.44\\
\hline
Heating oil & CME-US & 1998-2009 & up to 18 & 2835 & 0.55\\
\hline
Gasoil & ICE-Eu & 2000-2009 & up to 12 & 2546 & 0.45\\
\hline
Nat. gas (US) & CME-US & 1998-2009 & up to 36 & 3140 & 0.74\\
\hline 
Nat. gas (Eu) & ICE-Eu & 1997-2009 & up to 9 & 3055 & 0.63\\
\hline
\hline
Wheat & CME-US & 1998-2009 & up to 15 & 3026 & 0.88\\
\hline
Soy bean & CME-US & 1998-2009 & up to 14 & 2977  & 0.66\\
\hline
Soy oil & CME-US & 1998-2009 & up to 15 & 3056 & 0.81\\
\hline
\hline
Eurodollar & CME-US & 1997-2009 & up to 120 & 3056 & 0.64\\
\hline
Euribor & NYSE-Eu & 2000-2010 & up to 39 & 3036 & 0.63\\
\hline
Sterling futures & Euronext-Eu  & 1997-2010 & up to 36 & 3451 & 0.55\\
\hline
Gold & CME-US & 1998-2009 & up to 60 & 2877 & 0.99 \\
\hline
\end{tabular}
}
\end{table}
%%%%%%%%%%%%%%%%%%%%%%%%%%%%%%%%%%%%%%%%%%%%%%%%%%%%%%%%%%%%%%%%%%%%%%%%%%%%%%%%
For our empirical study, we selected $13$ futures markets corresponding to three classes of assets: two categories of commodities (energy and agricultural products) and financial assets. While in the majority of cases financial assets are interest rates on different currencies (Dollar, Euro and Sterling), they also include futures contracts on gold. The latter indeed is more an investment support than a consumption good. Among the different futures contracts negotiated worldwide, we retained those characterized by the largest transaction volumes and the longest maturities, over a long time period. We used the database Datastream in order to collect settlement prices on a daily basis. We rearranged the data in order to obtain time series with constant maturities. Lastly, we had to remove or merge some of the maturities in order to compare different markets on the same period. \\
Table $1$ presents our database: the different underlying assets of the futures contracts selected, the name of the futures exchange where transactions take place and its localization, the time period, the length of the prices curve, and the number of futures prices recorded. The last column provides a synthetic information on the degree of contango $C$ reported on each market during the period under examination. Contango (backwardation) corresponds to a situation where the deferred futures price is higher (lower) than the nearest price. \\
In derivative markets, temporal price relationships (that is to say contango and/or backwardation) result from arbitrage operations. The spread between the deferred and the near-term prices corresponds to the net carrying costs of the underlying asset of the contract, that is to say the difference between the costs and benefits of holding the contracts' support. The net carrying costs are usually positive for financial assets like interest rates: intuitively when the transactions' horizon is extended, it should become more and more expensive to borrow money.  As far as commodities are concerned, things are more contrasted. Some of the markets are most of the time in contango, as is the case for Wheat (Fig.\,\ref{ret_price} (a)); others, like for example Light crude oil (Fig.\,\ref{ret_price} (b)), are more often in backwardation. Such a situation reflects the fact that there is a premium for the immediate delivery of the commodity.  In this study, we measured the degree of contango of all markets by taking the difference between the $9$th month - this delivery date is the longest common maturity in our database - and the shortest available maturity. The indicator $C$ represents the fraction of records corresponding to a contango on the whole period. The figure reaches its maximal value for Gold, which is always in contango. As expected, the interest rates markets are more often in contango, and the same is true for all agricultural products and the two natural gases.
%%%%%%%%%%%%%%%%%%%%%%%%%%%%%%   Fig 1  %%%%%%%%%%%%%%%%%%%%%%%%%%%%%
%%%%%%%%%%%%%%%%%%%%%%%%%%%%%%%%%%%%%%%%%%%%%%%%%%%%%%%%%%%%%%%%%%%%%%
\section{Prices fluctuations}
In order to examine the statistical properties of price's fluctuations on the selected markets, we computed the prices returns $r(t)$ by taking the logarithm difference between two consecutive prices $P(t)$:
\begin{equation}\label{defret}
r\left(t\right)= \frac{\ln\left(P\left(t\right)\right) - \ln\left(P\left(t-\Delta t\right)\right)}{\Delta t},
\end{equation}
where $\Delta t= 1$ day, except during week-ends or days-off. In order to avoid bias in the statistics, returns are not computed when $\Delta t$ exceeds three days. Previous studies on prices returns in financial markets (\cite{Matteo2001}, \cite{Yamasaki2005}) alternatively used normalized or simple returns. After having checked that the results do not change with one or the other method, we retained the one defined by \ref{defret}. \\
The comparison of prices' returns for different delivery dates, illustrated by Figures \ref{ret_price}(c) and (d), shows that all time series have stochastic fluctuations around zero but also that the level of the fluctuations changes significantly with the maturity. This is a main feature of the term structures: the short maturities are affected by strong fluctuations while the long-term prices are less volatile. Thus the variance of the prices diminishes with the maturity. This decreasing pattern is usually referred to as the \emph{Samuelson effect} (\cite{Samuelson1965}). Intuitively, it happens because a shock affecting the nearby contract price has an impact on succeeding prices that decreases as maturity increases. Indeed, as futures contracts reach their expiration date, they react much stronger to information shocks, due to the ultimate convergence of futures prices to spot prices upon maturity. These price disturbances influencing mostly the short-term part of the curve are due to the spot market. \\
 Numerous works (\cite{Anderson1985}, \cite{Milonas1986} and \cite{Fama1987}) provided empirical support for this hypothesis for a large number of commodities and financial assets. In the case of commodities, in \cite{Chambers1996} and \cite{Deaton1992}, the authors observed that the Samuelson effect depends on the storage costs. More precisely, when the cost of storage is high, relatively little transmission of shocks via inventory occur across periods. Futures price's volatility consequently declines rapidly with the maturity. Moreover, there is a modified Samuelson effect in the case of seasonal commodities. Lastly, as far as the interest rates are concerned, the Samuelson effect can be in conflict with the monetary policy, especially on the shortest maturities. 
%%%%%%%%%%%%%%%%%%%%%%%%%%%%%%%%%%%%%%%%%%%%%%%%%%%%%%%%%%%%%%%%%%%%%%%%%%%%%%%%
\section{Statistical properties}\label{Stat_Prop}
%%%%%%%%%%%%%%%%%%%%%%%%%%%%%%%%%%%%%%%%%%%%%%%%%%%%%%%%%%%%%%%%%%%%%%%%%%%%%%%%
In this section we characterize the statistical properties of the stochastic processes underlying the returns of commodities and financial assets in the maturity dimension. We indeed compute the 1st to 4th moments and the tail of the distribution of the daily logarithm prices changes. 
%%%%%%%%%%%%%%%%%%%%%%%%%%%%%%%%%%%%%%%%%%%%%%%%%%%%%%%%%%%%%%%%%%%%%%
\subsection{Mean absolute returns} \label{meanmean}
%%%%%%%%%%%%%%%%%%%%%%%%%%%%%%%%%%%%%%%%%%%%%%%%%%%%%%%%%%%%%%%%%%%%%%
%%%%%%%%%%%%%%%%%%%%%%% Fig 2 %%%%%%%%%%%%%%%%%%%%%%%%%%%%%%%%%%%%%%%%
\begin{figure}[!t]
 \includegraphics[width=8.6cm,clip]{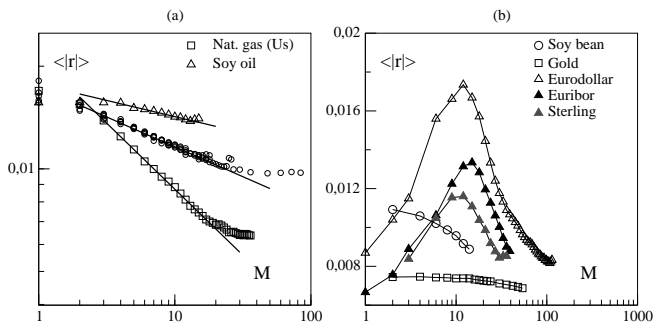}
\caption{Mean absolute returns as a function of the maturity. Left panel: Commodities following a power law function with exponents (from bottom to top) $\alpha_{mean}= 0.2 \ , \ 0.175 \ , \ 0.095$. The circles stand for all commodities having an exponent close to $0.175$. For the sake of simplicity all curves have been shifted to the same origin. The axes are in log scale. Right panel: Mean absolute returns for the futures contracts on interest rates, Gold, and Soy bean. The abscissa is in log scale.}
\label{av_abs}       
\end{figure}
%%%%%%%%%%%%%%%%%%%%%%%%%%%%%%%%%%%%%%%%%%%%%%%%%%%%%%%%%%%%%%%%%%%%%%
In order to examine the first moment of the distribution we compute the mean of the absolute daily returns. The latter is defined as follows: 
\begin{equation}
\left<|r|\right>_i = \frac{1}{T} \sum_{i=1}^{T}|r_i| , 
\end{equation}
where $T$ denotes the total number of records and $r_i$ the return at time $i$.\\
Figures \,\ref{av_abs}(a) and (b) reproduce the behavior of the mean absolute returns as a function of the maturity $M$. A decreasing pattern with the transactions' horizon is observed for commodities, which reflects the Samuelson effect. Among financial assets, Gold exhibits the flattest curve. Conversely, interest rates are characterized by the presence of a bell curve. The short-term fluctuations are lower than the mid-term ones: the monetary policy has a stabilizing influence on interest rates, and influences mainly the short-term part of the curve. It contradicts the Samuelson effect up to 12 months.Then, the decreasing pattern observed for commodities appears again (Fig. \,\ref{av_abs}(b)).\\
Another interesting result is that the fluctuations of commodity prices can be well described by a power law, as suggested by Figure \ref{av_abs}(a), except for Soy bean. The latter, as well as Gold, follows a linear relation with the maturity. Up to now, we did not identify why the Soy bean stands apart. As far as Gold is concerned, as previously mentioned, this asset does not really belong to the class of commodities.\\
Most of the commodities futures contracts under consideration have thus power law decreasing mean returns. Moreover, in their majority, the commodities follow a well defined scaling behavior $\left|r\right| \sim M^{-\alpha_{mean}}$ with a median value $\alpha_{mean}$ close to $0.175 \pm 0.012$. The American Natural gas follows another power law. In \cite{Lautier2010} the authors observed that futures prices in this market have a dynamics in the maturity space which is different from that observed in other markets. More precisely, the cross-correlations between the different maturities are subject to frequent and important destabilizations.\\
Finally, a crossover appears, after the $24$th month, on Figure \ref{av_abs}(a): at this point, the power law does not hold any more, and the mean fluctuations decrease much more slowly. This phenomenon is observed on two markets: the American Light crude oil and Natural gas. This crossover might result from the presence of preferred habitats (\cite{Modigliani1966}) for operators in commodity derivative markets, leading to different behavior of futures prices according to the range of maturity they belong to (\cite{Lautier2005}). \\
The presence of a power law for commodities can be interpreted as the signature of common underlying processes driving the dynamics of prices movements : temporal arbitrage between maturities and the Samuelson effect are two good candidates. 
 %%%%%%%%%%%%%%%%%%%%%%%%%%%%%%%%%%%%%%%%%%%%%%%%%%%%%%%%%%%%%%%%%%%%%%
\subsection{Variance}
%%%%%%%%%%%%%%%%%%%%%%%%%%%%%%%%%%%%%%%%%%%%%%%%%%%%%%%%%%%%%%%%%%%%%%
%%%%%%%%%%%%%%%%%%%%%%% Fig 3 %%%%%%%%%%%%%%%%%%%%%%%%%%%%%%%%%%%%%%%%
\begin{figure}[!t]
\includegraphics[width=8.6cm,clip]{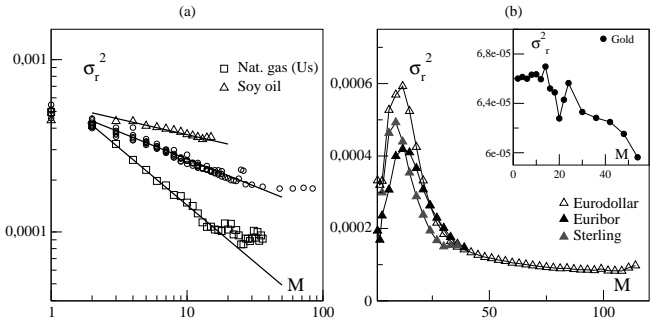}
\caption{Variance as a function of the maturity. Left panel: commodities following a power law function with exponents (from bottom to top) $\alpha_{var}=0.333 \ , \ 0.175 \ , \ 0.181$. The circles stand for all commodities having an exponent close to $0.175$. For the sake of simplicity all curves have been shifted to the same origin and the abscissa is in log scale. The axes are in log scale. Right panel: futures contracts on financial assets.}
\label{var_ret}       
\end{figure}
%%%%%%%%%%%%%%%%%%%%%%%%%%%%%%%%%%%%%%%%%%%%%%%%%%%%%%%%%%%%%%%%%%%%%%
The analysis of the variance of the daily returns fluctuations, $\sigma^2_r$, reinforces the conclusions reached with mean returns. We compute the variance as follows: 
\begin{equation} 
\sigma^2_r = \frac{1}{T}\sum_{i=1}^{T}\left(r_i - <r>_i\right)^2
\end{equation}
Figures \ref{var_ret}(a) and (b) present our results. As was the case for the mean returns, we observe a decreasing pattern of the fluctuations, reminiscent of the Samuelson effect. Furthermore all commodities, including Soy bean, can now be described by a power law. Table 2 exhibits the values of the exponents of the power laws obtained for commodities, for the mean absolute returns and the variances as well as the errors  $\Delta$ on these measures. Whatever the commodity is concerned, the latter are low. Now the distinction between the financial underlying assets and the commodities is very clear. Each of these category exhibits homogeneous behavior. 
%%%%%%%%%%%%%%%%%%%%%%% TABLEAU 2 %%%%%%%%%%%%%%%%%%%%%%%%%%%%%%%%%%%%
\begin{table}[!b]
\caption{Exponents of the power law function for the mean and variance of the returns.\\}
\label{tab_vol_sigma}
\resizebox{0.4\textwidth}{!}{
\begin{tabular}{|l||l|l||l|l|}
\hline
Futures & $\alpha_{mean}$ & $\Delta \alpha_{mean}$ & $\alpha_{\sigma^2}$ & $\Delta \alpha_{\sigma^2}$ \\
\hline
Soy oil & 0.095 & 0.004 & 0.181 & 0.008 \\
Soy bean & no & no & 0.267 & 0.017  \\
Wheat & 0.198 & 0.007 & 0.332 & 0.021 \\
Light crude & 0.179 & 0.001 & 0.362 & 0.002\\
Brent crude & 0.160 & 0.002 & 0.315 & 0.003 \\
Heating oil & 0.188 & 0.004 & 0.353 & 0.013 \\
Gasoil & 0.163 & 0.003 & 0.321 & 0.002\\
Nat. gas (Eu) & 0.2 & 0.005 & 0.333 & 0.02\\
Nat. gas (Us) & 0.387 & 0.005 & 0.664 & 0.022\\
\hline
\end{tabular}
}
\end{table}
%%%%%%%%%%%%%%%%%%%%%%%%%%%%%%%%%%%%%%%%%%%%%%%%%%%%%%%%%%%%%%%%%%%%%%
%%%%%%%%%%%%%%%%%%%%%%%%%%%%%%%%%%%%%%%%%%%%%%%%%%%%%%%%%%%%%%%%%%%%%%
\subsection{Skewness} \label{skew}
%%%%%%%%%%%%%%%%%%%%%%%%%%%%%%%%%%%%%%%%%%%%%%%%%%%%%%%%%%%%%%%%%%%%%%
%%%%%%%%%%%%%%%%%%%%% Fig 4 %%%%%%%%%%%%%%%%%%%%%%%%%%%%%%%%%%%%%%%%%%
\begin{figure}
  \includegraphics[width=8.6cm,clip]{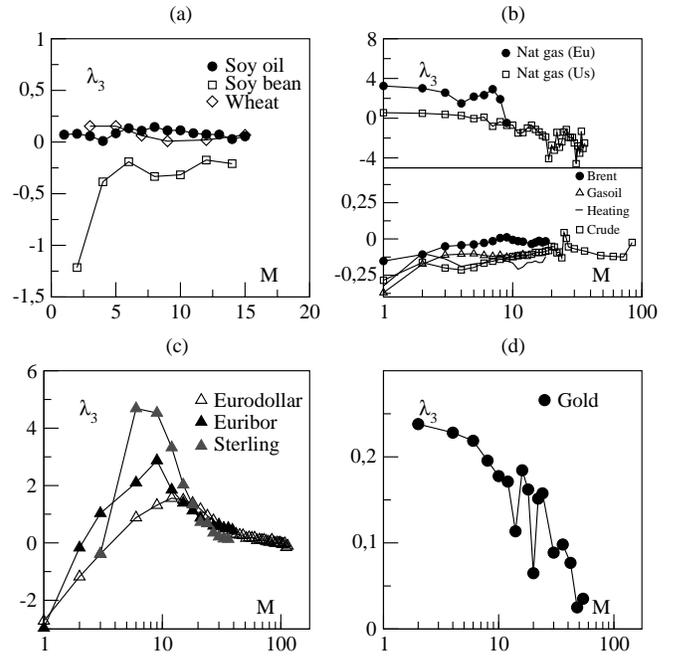}
\caption{Skewness as a function of the maturity. (a) agricultural products; (b) energy sector; (c) interest rates; (d) gold. The abscissae of figures (b), (c) and (d) are in log scale}
\label{skew_ret}       
\end{figure}
%%%%%%%%%%%%%%%%%%%%%%%%%%%%%%%%%%%%%%%%%%%%%%%%%%%%%%%%%%%%%%%%%%%%%%
Let us now turn to the third moment of the distribution. We compute the skewness $\lambda_3$ of the returns as follows:
\begin{equation}
\lambda_3 = \frac{1}{T}\sum_{i=1}^{T}\frac{\left(r_i-<r>_i\right)^3}{\sigma_r^3}
\end{equation}
This measure gives the level of asymmetry of the probability distribution of a random variable. A negative (positive) skewness indicates that the values are distributed to the right (left) of the mean.\\
Figure \,\ref{skew_ret} provides the results for agricultural products, energy products, interest rates and Gold. Interest rates exhibit a quite homogeneous behavior, with a negative skewness for the shortest maturities, which turns into a positive one for maturities around one year, and then a tendency toward zero. Thus, fluctuations are usually high for the short maturities, low for the middle ones, whereas the distribution becomes symmetrical for longer maturities. As far as the other assets are concerned, the behavior of the skewness with the maturity is generally more regular: it is positive and decreases with the maturity for Soy oil, Soy bean, the two natural gases and Gold. Conversely it is negative and increases with the delivery dates for the group of petroleum products. Thus, the products characterized by a very frequent contango seem to exhibit positive skewness, whereas backwardated markets appear to be associated with negative skewness. \\
Such a result is consistent with the fact that prices' fluctuations are not the same in contango and backwardation, especially in commodity markets. Such markets indeed are characterized by a positive constraint on inventory, which does not hold for financial assets used for investment purposes.  When stocks are rare, in backwardation, arbitrage operations are all the more unlikely to happen than the shortage is pronounced. In such a case, the level of prices' spread is solely determined by the spot price the operators are willing to pay in order to immediately obtain the merchandise. Moreover, because inventories are not sufficiently abundant to absorb the fluctuations in the demand, the spot price is volatile, and so is the prices' spread. Thus a longer left tail for the distribution, especially for shorter maturities, is not a surprise. The positivity constraint disappears in contango, when stocks are abundant. In such a case, prices' spread are stable and, under the pressure of arbitrage operations, they are limited to the level of storage costs. A positive skewness is thus probable.\\
Lastly, compared with the other assets, Gold exhibits a specific behavior: the skewness is positive, and decreasing, which could have been expected for a market which is almost always in contango.
%%%%%%%%%%%%%%%%%%%%%%%%%%%%%%%%%%%%%%%%%%%%%%%%%%%%%%%%%%%%%%%%%%%%%%
\subsection{Kurtosis}\label{kurto}
\begin{figure}
  \includegraphics[width=8.6cm,clip]{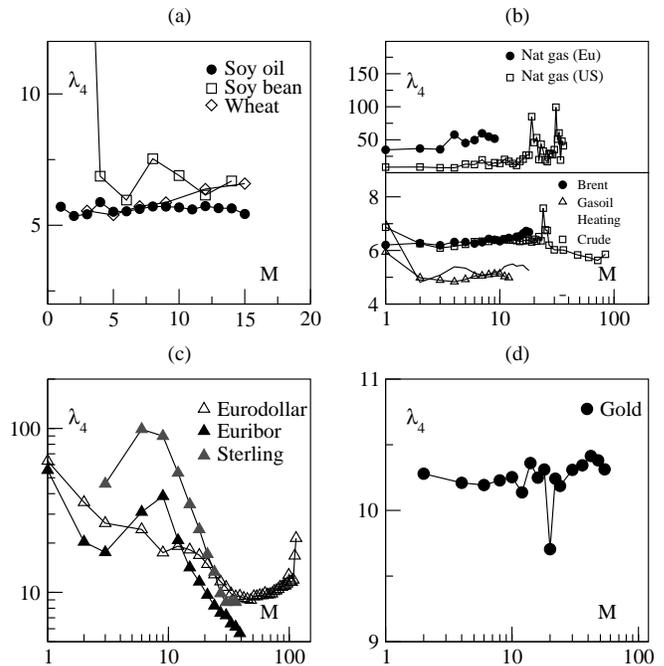}

\caption{Kurtosis as a function of the maturity. (a) agricultural products; (b) energy sector; (c) interest rates; (d) Gold. The abscissae of figures (b), (c) and (d) are in log scale.}
\label{kurt_ret}       
\end{figure}

%%%%%%%%%%%%%%%%%%%%%%%%%%%%%%%%%%%%%%%%%%%%%%%%%%%%%%%%%%%%%%%%%%%%%%
The kurtosis of the distribution $\lambda_4$ was computed in the following way: 
\begin{equation}
\lambda_4 = \frac{1}{T}\sum_{i=1}^{T}\frac{\left(r_i-<r>_i\right)^4}{\sigma_r^4}
\end{equation}
Figure \,\ref{kurt_ret} displays our results. They are in line with our comments on skewness, and more precisely with the observation that there is a quite homogeneous behavior of the distribution among one class of assets. Moreover, the fourth moment of the distribution is generally high, whatever the asset is considered. The degree of peakedness of the distribution is however especially important for the two natural gases and the short-term maturities of interest rates. Thus for these markets, a large part of the return's variance is due to infrequent extreme deviations. \\
As far as interest rates are concerned, the presence of few large deviations in the returns is consistent with isolated actions of the  
monetary authorities. As for natural gases, while the high kurtosis on short-term maturities is probably due to storage difficulties, such an explanation does not hold for the long-term maturities on the American gas. In this case, the lack of stability of this market, previously mentioned, can be invoked.
%%%%%%%%%%%%%%%%% Fig 5 %%%%%%%%%%%%%%%%%%%%%%%%%%%%%%%%%%%%%%%%%%%%%%
%%%%%%%%%%%%%%%%%%%%%%%%%%%%%%%%%%%%%%%%%%%%%%%%%%%%%%%%%%%%%%%%%%%%%%
%%%%%%%%%%%%%%%%%%%%%%%%%%%%%%%%%%%%%%%%%%%%%%%%%%%%%%%%%%%%%%%%%%%%%%
\section{Tail exponent term structure}\label{tailtermstruc}
%%%%%%%%%%%%%%%%%%%%%%%%%%%%%%%%%%%%%%%%%%%%%%%%%%%%%%%%%%%%%%%%%%%%%%
In this section we address the question of whether the scaling properties of returns probability distributions change with the maturity: in other words, is there a term structure of tail exponents for derivatives?  As mentioned previously, if stocks and foreign exchange markets have received a lot of attention, such was not the case for commodities. Moreover, except for interest rates, the maturity dimension has been omitted (\cite{Caju2009}, \cite{Caju2007}, \cite{Tabak2005}, \cite{Caju2007_2}, \cite{Matteo2001}).\\
One of the most frequent empirical findings concerning price fluctuations of assets is the inverse cubic law, which stipulates that the tails of the returns are power law distributed with an exponent $\mu + 1 \sim 4$. This behavior seems to be \emph{universal}. It was observed on several financial markets (stocks, stock indexes, exchange rates, interest rates, and the nearest delivery dates of commodities), different time scales (investigations where carried on time intervals varying from minutes to months) and different time periods (\cite{Gopi1999}, \cite{Lux1996}, \cite{Gopi1998}). \\ 
In more specific studies, several authors observed that the tail exponent remains outside the L\'evy stable domain, within a range of $3$ to $5$, for symmetric as well as for asymmetric tails (\cite{Cont2001}). As also shown in \cite{Bouchaudbook}, the estimate of the exponent can be sensitive to the time scale and $\mu$ is lower for high frequency data, compared to the figures obtained with weekly or monthly time series. Even if, on the basis of empirical data, a precise determination of the tails remains hard, finding an accurate value for the exponent is an important issue. More precisely, the finite fourth normalized cumulant requires $\mu > 4$, otherwise the kurtosis is ill-defined and may lead to tricky conclusions.\\ 
The literature provides several methods for the estimation of the tail exponent. Many works resort to the Hill estimator, which is the conditional maximum likelihood estimator for a pure power law distribution, and is based on the $k$ largest order statistics (\cite{Hill}). The easy computation and the accuracy of this estimator, at least for some statistical distributions, made it very famous. It is the one used in \cite{Matia_pre_2002} in order to distinguish between the scaling properties of stocks and commodities. In \cite{Plerou_pre_2007}, the authors introduce another estimator, the so-called MS estimator (\cite{MS1998}) and compare its accuracy with the previous one. They show that the tail exponent of certain markets strongly depends on the estimation threshold retained and that close to the limit of the stable L\'evy regime, the estimator is not reliable. It is thus important to retain a method which is not dependent of any threshold.\\
In our study, we retained  the procedure described in \cite{Clauset2009} which, first does not require any threshold and second, uses maximum-likelihood fitting methods with goodness-of-fit tests. The latter are based on the Kolmogorov Smirnov statistic and on likelihood ratios. Finally, in order to obtain the most accurate values for the tail exponent and standard errors, we combined their method with bootstraped samples. \\
In what follows, we first present the results obtained for each market. We then propose a more general analysis, based on averaged tail exponents.
%%%%%%%%%%%%%%%%%%%%%%%%%%%%%%%%%%%%%%%%%%%%%%%%%%%%%%%%%%%%%%%%%%%%%%
\subsection{Overview of the results for each market}\label{Markover}
%%%%%%%%%%%%%%%%%%%%%%%%%%%%%%%%%%%%%%%%%%%%%%%%%%%%%%%%%%%%%%%%%%%%%%
%%%%%%%%%%%%%%% Fig 6 %%%%%%%%%%%%%%%%%%%%%%%%%%%%%%%%%%%%%%%%%%%%%%%%

\begin{figure}
  \includegraphics[width=8.6cm,clip]{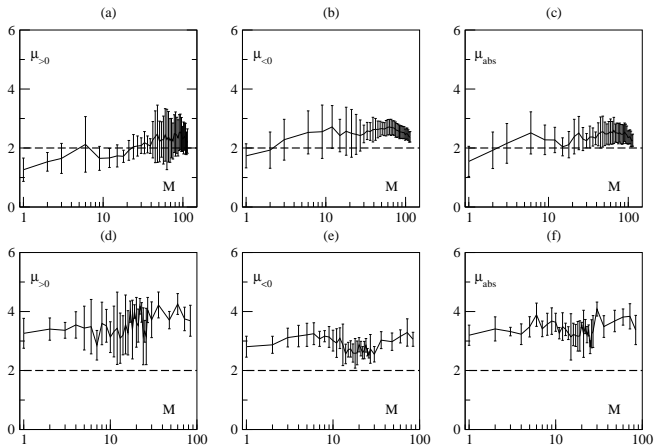}

\caption{Tail exponent term structures. Upper panel: Eurodollar; Lower panel: Light crude oil. From the left to the right:  Positive tail; Negative tail; Absolute returns. The dashed line corresponds to the limit of the L\'evy stable regime. The abscissae are in log scale.}
\label{muIEDNCL}       
\end{figure}
\begin{figure}
  \includegraphics[width=8.6cm,clip]{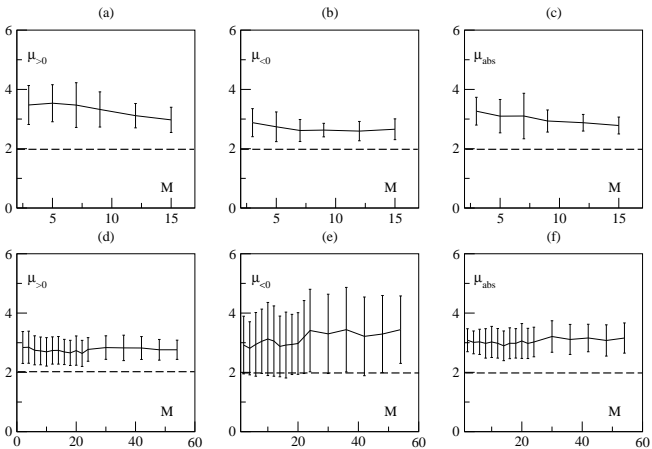}

\caption{Tail exponent term structures. Upper panel: Wheat; Lower panel: Gold. From the left to the right : Positive tail; Negative tail;  Absolute returns.}
\label{muCWNGC}       
\end{figure}

%%%%%%%%%%%%%%%%%%%%%%%%%%%%%%%%%%%%%%%%%%%%%%%%%%%%%%%%%%%%%%%%%%%%%%
The study of the term structures of the tail exponents for the $13$ futures contracts under examination leads us to several results, as illustrated by Figures \,\ref{muIEDNCL} and  \,\ref{muCWNGC}. The latter provide representative examples of the positive, negative and absolute exponent term structures obtained for each category of futures contracts, that is to say Eurodollar, Light crude oil, Wheat and Gold.\\
First, most of the returns do not belong to the L\'evy stable domain, whatever the maturity is considered. Thus, in their majority, the distributions of the returns cannot be described, neither by stochastic processes with stable laws and infinite variance, nor by brownian motions. The exceptions are the European Natural gas and the first maturities of the three interest rates contracts. Due to monetary policy actions, governments indeed often maintain the same level of interest rates during several months. Over reaction of the traders to sudden changes in the level of politically driven interest rates might explain the greater probability of high extreme events. \\
Second result, the distribution of absolute returns exhibits an increasing term structure of the tail exponent for Light crude oil, Gold, Heating oil and the three interest rates. We observed the opposite behavior for Wheat and the two natural gases. No specific tendency can be found on the agricultural products.\\
Third result, some of the futures contracts, that is to say Light crude oil and Gold, are characterized by a strong asymmetry between the positive and negative parts of the distribution (Fig. \,\ref{muIEDNCL}(d,e) and Fig. \,\ref{muCWNGC}(d,e)). Oil is distinguishable as it exhibits relatively few rare events on its right tail. The same phenomenon is typical of the left tail of the Gold contract. As a consequence, these two contracts exhibit relative high errors and exponents' level on these sides of their distribution. As mentioned in \ref{skew}, this might be attributed to the level of contango and / or backwardation of these markets.\\
Lastly, the interest rates contracts share common patterns: the exponents of the short-term maturities belong to the L\'evy stable domain. Moreover, $\mu$ increases slowly with the maturity, thus indicating a damping of extreme price movements, and reaches a plateau at long time scale. The same kind of conclusion has been reached by the authors in \cite{Matteo2001} on the Eurodollar. In their study, they compare the probability distribution with the general class of L\'evy, Khinchtine stable distributions. They observe that from $1990$ to $1996$ the tail region is in the L\'evy stable domain. They also expect a faster decrease for larger fluctuations as would be the case for truncated L\'evy flights \cite{Mantegna1994}. We thus find, on a latter period ($1998-2010$), similar values for the short-time part of the prices curve as well as a faster vanishing for higher maturities. Moreover, we generalize these results on other interest rates. 
%%%%%%%%%%%%%%% Fig 7 %%%%%%%%%%%%%%%%%%%%%%%%%%%%%%%%%%%%%%%%%%%%%%%%
%%%%%%%%%%%%%%%%%%%%%%%%%%%%%%%%%%%%%%%%%%%%%%%%%%%%%%%%%%%%%%%%%%%%%%

%%%%%%%%%%%%%%%%%%%%%%%%%%%%%%%%%%%%%%%%%%%%%%%%%%%%%%%%%%%%%%%%%%%%%%
\subsection{Generalized exponents term structure}\label{agregatedata}
\begin{figure}
  \includegraphics[width=8.6cm,clip]{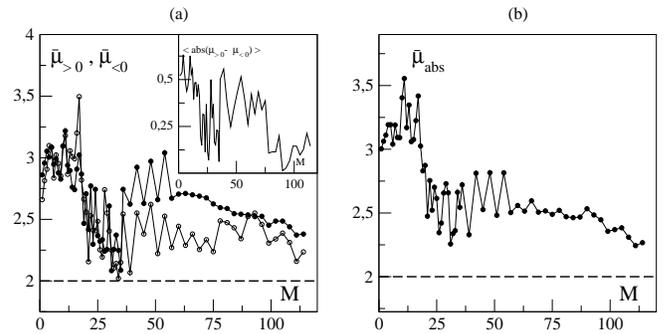}

\caption{Aggregate tail exponents term structures; (a): Positive and negative aggregate tail exponents; Inset: distance between the positive and negative aggregate tail exponents; (b) Absolute returns}
\label{agregate}       
\end{figure}

%%%%%%%%%%%%%%%%%%%%%%%%%%%%%%%%%%%%%%%%%%%%%%%%%%%%%%%%%%%%%%%%%%%%%%
In this section, we extend in the maturity dimension the results of \cite{Matia_pre_2002}. The authors indeed compare the scaling properties of spot and futures prices of commodities. They however do not precise what kind of futures prices they use: our guess is that they retain the nearest available maturity. They compute average exponents for all markets under scrutiny and find $\bar{\mu}_{\mathrm{spot}}\sim 2.3$ and $\bar{\mu}_{\mathrm{futures}}\sim 3.2$.
 As far as our study is concerned, as we aim to give a deeper insight of the maturity dimension, we calculate, for each maturity $M$, the average, positive, negative and absolute tail exponents:
\begin{equation}
\bar{\mu}\left(M\right)=\frac{1}{N\left(M\right)}\sum_{i=1}^{N\left(M\right)}\mu_i\left(M\right),
\end{equation}
where $\bar{\mu}\left(M\right)$ can be computed for absolute, positive, negative tails and $N\left(M\right)$ is the number of futures contracts having maturity $M$.\\ 
We thus test whether the inverse cubic law can be observed in the maturity dimension. If this is true, this would suggest the presence of identical trading behavior for assets traded on the spot and derivative markets.\\
 We present the results on the Figure \ref{agregate}. The average positive and negative exponents' curves roughly collapse from the first to the thirty-six months with a minimum close to the L\'evy stable region. Then they separate from each other and the values of $\bar{\mu}$ become greater for left tails. The degree of asymmetry between the two tails is measured by the distance $|\mu_{>0}-\mu_{<0}|$ (inset of Figure \ref{agregate}(a). We observe a segmentation along the term structure into four parts: from the first to the eighteenth month, the figures are close to $0.5$; then, until the 35 months' maturity, they decrease around $0.2$. From $36$ to $75$ months they reach a plateau, close to $0.4$. Finally, the curve goes down and stabilizes around $0.1$. Thus the presence of preferred habitats, usually detected in prices (\ref{meanmean},\cite{Lautier2005}, \cite{Gabillon1995}), can also be observed in the asymmetry of extreme events.\\
We finally studied the behavior of absolute returns, as displayed by Figure \ref{agregate}(b) and observe two regimes of values. A first plateau is located around $\bar{\mu}_{abs}\sim 3.15$ and is surrounded by the $1st$ and $18th$ months. A second one starts at $19$ months and ends at the latest maturities, for a value close to $\bar{\mu}_{abs}\sim 2.53$. In \,\cite{Matia_pre_2002}, the authors find a futures power law decay with an exponent close to $3.2$, which is in good agreement with our values, for the first part of the curve. As the authors did not give any information about the maturity of the futures prices used, we cannot precisely localize their value on the first plateau.\\
The existence of these two regimes for the tail exponent reminds of a phase diagram. This result suggests that in the idea of a thermodynamic limit, with an infinite number of markets, the curve $\bar{\mu}_{abs}\left(M\right)$ could be defined as follows:
\begin{equation}
\bar{\mu}_{abs}\left(M\right)\sim
\left\lbrace\begin{array}{l}
3  \ \mathrm{if} \ M<M_t,\\
2.5  \ \mathrm{if} \ M>M_t,
\end{array}
\right.
\end{equation}  
with $M_t=18$ months. This transition value $M_t$ marks off two regions of extreme events. The first is close to the spot price and the shock regime is probably mostly originated from the physical market and inventories. The second region is characterized by another type of shocks which might be due to a lack of liquidity or financial activity. In such a perspective, we hypothesize that the maturity $M_t$ defines a time horizon delimiting two regimes of risk. 
%%%%%%%%%%%%%%% Fig 8 %%%%%%%%%%%%%%%%%%%%%%%%%%%%%%%%%%%%%%%%%%%%%%%%
%%%%%%%%%%%%%%%%%%%%%%%%%%%%%%%%%%%%%%%%%%%%%%%%%%%%%%%%%%%%%%%%%%%%%%

%%%%%%%%%%%%%%%%%%%%%%%%%%%%%%%%%%%%%%%%%%%%%%%%%%%%%%%%%%%%%%%%%%%%%%%%%%%%%%%%
\section{Conclusions}\label{Conclu}
%%%%%%%%%%%%%%%%%%%%%%%%%%%%%%%%%%%%%%%%%%%%%%%%%%%%%%%%%%%%%%%%%%%%%%%%%%%%%%%%
In this article we examine and compare the behavior of the term structure of futures prices' returns of commodity and financial derivatives. Compared to financial assets, commodities have received less attention; furthermore, the temporal aspect of the term structure has often been discarded. Whereas the first and second moments of the distribution exhibit a bell curve for interest rates, with a maximum located at a maturity indicating a limit of the monetary policy influence, the commodities can be distinguished by a decreasing pattern with the maturity. This phenomenon is usually referred to as the \emph{Samuelson effect} and can be characterized by a well-defined exponent for most of the commodities under examination.\\
Lastly, the analysis of the skewness and kurtosis shows that derivative markets tend to exhibit an asymmetrical distribution, skewed to the left (right) when they are in contango (backwardation). All these results lead us to conclude that the term structure of the first fourth moments of the distributions provide statiscal signatures for commodities and interest rates.\\
Finally, the study of rare events shows an new feature of derivative markets. The value of the average tail exponent defines a phase diagram with two phases separated at the maturity of transition $M_t=18$ months, reflecting a segmentation in the maturity dimension. Each of these phases corresponds to a specific regime of risk.\\
Further developments in the modeling of commodities prices and interest rates should take into account these new empirical facts. Moreover, future research in this field will have to highlight the origin of microscopic and/or macroscopic forces responsible for such statistical properties.
%%%%%%%%%%%%%%%%%%%%%%%    Bibliography   %%%%%%%%%%%%%%%%%%%%%%%%%%%%

%%%%%%%%%%%%%%%%%%%%%%%%%%%%%%%%%%%%%%%%%%%%%%%%%%%%%%%%%%%%%%%%%%%%%%

%\bibliographystyle{apsrmp}
%\bibliographystyle{unsrt}
%\bibliographystyle{h-physrev3}
%\bibliography{divers,biologie,biophys,collective,statphys}

\end{document}